\documentclass{article}
\usepackage[utf8]{inputenc}
\usepackage{amsmath}
\usepackage{graphicx}
\usepackage{hyperref}

\title{Simulation tumor growth in heterogeneous medium based on diffusion equation}
\author{Maxim V. Polyakov$^*$ and Valeria V. Ten\footnote{Volgograd State University, Universitetsky pr., 100, Volgograd, Russia}}
\date{February 06, 2023}

\begin{document}

\maketitle

\begin{abstract}
In the present article the diffusion equation is used to model the spatio-temporal dynamics of a tumor, taking into account the heterogeneous of the medium. This approach makes it possible to take into account the complex geometric shape of the tumor in modeling. The main purpose of the work is demonstration the applicability of this approach by comparing the obtained results with the data of clinical observations. We used an algorithm based on an explicit finite-difference approximation of differential operators for solve the diffusion equation. The ranges of possible values that the input parameters of the model can take to match the results of clinical observations are obtained. It is concluded that the heterogeneity of the physical parameters of the model, in particular the diffusion coefficient, has a significant effect on shape of tumor.

\vspace{2mm}
\end{abstract}

\section{Introduction} \label{intro}

\par 
The incidence of cancer according to the World Health Organization (WOS) is the second most common reason of death in the world, as it accounts for approximately 10 million deaths, or one in six deaths \cite{Ferlay}. Tumors arise due to the deformation of genes as a result of damage to the DNA of the body under the influence of various factors inherited by the body or acquired in the course of life. Benign tumors are a group of pathological cells limited by a certain cavity \cite{Liu2022,Kok2019}. They become malignant when the area of the tumor lesion extends beyond the cavity, going beyond the basement membrane. For the vital activity of pathological cells, a greater amount of nutrients and oxygen is required than in healthy cells, which is provided by their transportation through blood vessels. The phenomenon of angiogenesis is based on the formation of new blood vessels around cancer cells \cite{Frederique, Carmeliet}. In addition, metastasis is a significant problem. With the transfer of blood or lymph, malignant cells form a lesion in a new area of the body. Metastases are one of the main causes of cancer mortality \cite{Huysentruyt}. Tumor diseases are classified by the type of cells involved in the disease process and by the tissues in which they are formed. Currently, there are various approaches to cancer treatment. For example, radiation therapy, hormone therapy, tumor surgery, chemotherapy, immunotherapy, etc. \cite{Chabner, Demaria, Shepard, Morch, Farkona}. Each of these methods has its disadvantages, advantages and limits of applicability (including depending on the type of tumor itself). Moreover, the evaluation of the effectiveness of treatment by one of the methods or a combination of them is carried out for each individual neoplasm, taking into account a set of individual qualities of the organism that affect the tumor. With the passage of time and the development of the disease, the possible effectiveness and applicability of the chosen therapy naturally decreases. As cell populations accumulate more and more changes, they acquire characteristics that allow them to persist in biological tissues for a long time \cite{Armitage, Weinberg}. Often, with unsuccessful attempts to stop the spread of the tumor, when treatment becomes a complex process with the onset of the last stages, the moment of its uncontrolled growth begins, in which the probability of recovery is extremely small and rapidly decreases every day. The natural resistance of the organism to the disease in question is completely insufficient to slow down the effect of cancer cells on healthy cells and their spread. In addition, it is quite problematic to detect a tumor in the early stages. Therefore, this disease is considered one of the most dangerous.

\par To solve some of the above problems, methods of mathematical and numerical modeling are widely used, which consist in simulating growth at all stages of its development, when it is possible to predict how accurately the development of therapy will be and to identify a neoplasm in the disease. An important aspect is the prediction of the spread of pathological processes in cells in biological tissues, as well as stopping or slowing down this process. One of the works on the statistics of statistics is \cite{Hahnfeldt}. It predicts the dynamics of growth and endothelium. In \cite{Gandolfi} the proposed model is expanded and analyzed in a wider range of applications. finding that angiogenic therapy, as one of the research methods, is sufficient. Model-based studies have been encountered to derive composite structures and processes in cancer \cite{Anderson1,Byrne} and have been used to propose new experiments to develop different detection methods as well as disease risk analysis \cite{Lenaerts,Jones}. Quantitative descriptions of the mechanisms that lead to cancer at different scales of space and time lead to new problems, which are solved by conducting new experiments and applying mathematical models based on empirical data. In the work \cite{LIU20161} the identification of a periodic mathematical model of cancer treatment by radiation therapy with the identification of chemo-therapy. The obtained solutions make it possible to find conditions under which successful cancer treatment is possible.

\par At present, general mathematical models of tumors consider and take into account the structural nature of the tissue and the need of a growing tumor to involve blood vessels, due to which, from a certain moment, the neoplasm spreads. There are several groups of mathematical models for describing tumor diseases. There are multiphase models based on the fact that conventionally a tumor consists of at least three components: cells, extracellular matrix, and extracellular fluid. In such cases, the structure of the tumor is often simplified, considering it as a liquid, less often a solid substance \cite{Breward}. There are also nested models, the essence of which is to study the system at the cellular or tissue level. Here much attention is paid to the chemical characteristics of the considered medium \cite{Anderson2}. A model based on the dynamic interaction of a large number of cells should take into account, among other things, the physical properties of individual cells (volume, properties of various mechanical effects, etc.). Ordinary differential equations and partial differential equations \cite{BULAI2023721} are used together to model the growth of a metastatic tumor. In particular, such models are used to study lung cancer and breast cancer. Finally, hybrid models are used that combine the paradigms of both the discrete structure of the tumor and the continuum \cite{Drasdo1, Drasdo2}.

\par
\begin{figure}[ht] 
  \center
  \includegraphics [width = 0.65 \linewidth] {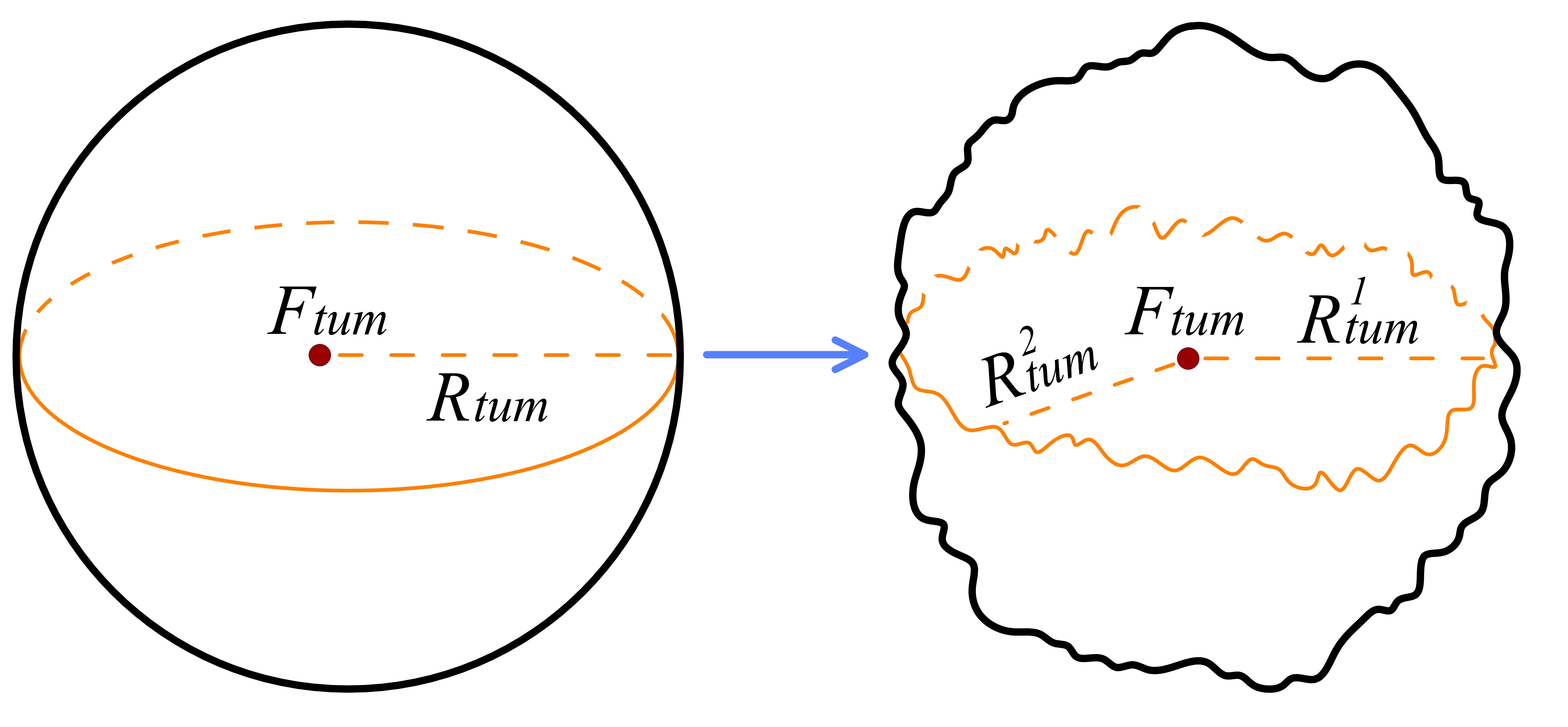}
  \caption{Approximations of the geometry of the cancer tumor model: a spherical tumor model of radius $R_{tum}$ with a focus of occurrence at point $F_{tum}$ (left), a spatially inhomogeneous tumor model, where $R^1_{tum}\neq R^2_{tum}$ with a focus of occurrence at point $F_{tum}$ (right).} 
  \label{img:heterogeneoustumor}  
\end{figure}

 In many works (see for example \cite{Hahnfeldt,Mattei2016AMM}), the mathematical model of tumor dynamics is based on systems of ordinary differential equations, where the volume of healthy tissue, the volume of tissue affected by a cancerous tumor and some others are used as calculated quantities. One of the assumptions of this approach is to take into account the tumor as a spherical object, although in reality the form of cancerous tumors has a much more complex spatial structure.
One of the goals of developing our model is the transition from a spherical tumor model to a spatially heterogeneous one (Fig. \ref{img:heterogeneoustumor}).

\par In the work \cite{SunJames2020}, a mathematical model of cancer development is proposed, according to which a growing tumor is described as a poroelastic medium consisting of solid and liquid components. In this mechanical description of the dynamics of tumor growth, the basic equations of tumor growth are introduced and combined with deformations and substantially nonlinear factors determining the physical process in order to increase the accuracy and efficiency of the modeling process. The study of the role of the internal heterogeneity of the cells of the body, arising from variations in the parameters of the cell cycle and apoptosis, is an important step towards better information about the drugs used. Such models \cite{ModelingCancerCell2017} allow taking into account an antimitotic drug whose effect on cell dynamics is to cause mitosis to stop, increasing the average duration of the cell cycle. A special place in the study of this problem is occupied by models of tumor growth based on cellular automata. Based on the data of mathematical models in \cite{SimulatedBainTumor2000}, the process of growth of a malignant brain tumor is investigated. It has been shown that the macroscopic behavior of a tumor can be realistically modeled using microscopic parameters. Based on such models, it is possible to predict the composition and dynamics of the tumor at specified points in time, which is sufficiently consistent with medical data. The article \cite{WEI2023645} offers a mathematical model of triple negative breast cancer. The mathematical model includes the interaction between tumor cells, innate and adaptive immune cells, programmed cell death protein, programmed death ligand and immune checkpoint inhibitor.

\par We note separately a class of inverse problems that allow us to determine the free parameters of the model by comparing the observed data and the simulation results. For example, in the work \cite{Polyakov}, the coefficient of thermal conductivity and the specific heat capacity of biological tissues, including tumor tissue, are calculated by optimization methods based on known distributions of breast temperature. A similar approach is also applicable for calculating the parameters of a cancerous tumor, according to the known dependencies of the dynamics of diameter and volume obtained on the basis of medical measurements.

\par In this article, we propose a method for modeling the spatial dynamics of a cancerous tumor based on the diffusion equation. This approach is applicable to the description of porous media, which is the system under study \cite{Galaguz}. Software has been developed for calculating the spatio-temporal dynamics of cancerous tumors based on grid numerical methods. We investigate the problem of selecting coefficients for the diffusion mass transfer equation describing the process of cancer tumor growth based on numerical modeling methods and using the results of medical examinations. In addition, ranges of values of the free parameters of the model satisfying the observed data are obtained. The analysis of the results confirms the applicability of the approach proposed in the work for modeling tumor growth.

\par Prefixing the presentation of our research results in Section \ref{intro}, we give a brief overview of the achievements in the study of the dynamics of cancer tumors both in medicine in general and in the field of mathematical modeling. 
Section \ref{MathematicalAndNumericalModel} describes a mathematical model, numerical methods and algorithms for solving the resulting dynamic system. In Section \ref{NumSim}, we present the results of our set of simulations and the degree to which they correspond to the observed data, and also give the limits of the applicability of the model. We summarize and discuss our results in Section \ref{concldiss}.

\section{Mathematical model and methods for numerical analysis of growth process of a cancerous tumor}\label{MathematicalAndNumericalModel}

%\subsection{Математическая модель пространственно-временной динамики раковой опухоли}
\subsection{Mathematical model of spatio-temporal dynamics of a cancerous tumor}
% В качестве базвовой модели для описания пространственной динамики раковой опухоли будем использовать уравнение диффузии, так как оно позволяет учесть нелинейность пространственного распространения и постепенное поражение новых областей еще здоровых клеток \cite{Aristov, Farre}. 
As a basic model for describing spatial dynamics of a cancerous tumor, we will use diffusion equation, since it allows us to take into account the nonlinearity of spatial distribution and the gradual damage to new areas of still healthy cells \cite{Aristov, Farre}. 
%Направление распространения будем задавать постоянным коэффициентом. 
The direction of propagation is given by a constant coefficient.
%Опишем процесс распространения раковых клеток в здоровых (нормальных) тканях с заданной скоростью в трехмерной системе координат
The process of spreading cancer cells in healthy tissues at a given rate in a three-dimensional coordinate system is described by the equation
\begin{eqnarray}
    \label{eq:equation2.1}
    \frac{\partial D}{\partial t} + \frac{\partial}{\partial x} u\cdot D + \frac{\partial}{\partial y} v\cdot D + \frac{\partial}{\partial z} w\cdot D
    =\nonumber\\ 
    \frac{\partial }{\partial x}P_{x}\frac{\partial D}{\partial x} + \frac{\partial }{\partial y}P_{y}\frac{\partial D}{\partial y} + \frac{\partial }{\partial z}P_{z}\frac{\partial D}{\partial z} + Q(x, y, z, t) - \alpha D,
\end{eqnarray}
where $u\equiv S_{x}$, $v\equiv S_{y}$, $w\equiv S_{z}$ is rate of spread of a cancerous tumor in space along corresponding axes of the Cartesian coordinate system, $D$ is degree of tissue damage in some area, $P$ is propagation factor, $Q$ is function that characterizes focus of disease $F_{tum}$, $\alpha$ is resistance parameter of the organism to tumor development.
The value $\alpha$ specifies the dependence of the system on use of any therapy, or the natural resistance of the disease of the immune system, or clearance.
%where $u\equiv S_{x}$, $v\equiv S_{y}$, $w\equiv S_{z}$ is скорость распространения раковой опухоли в пространстве по соответствующим осям декартовой системы координат, $D$ -- степень поражения ткани в некоторой области, $P$ --- коэффициент распространения, $Q$ -- функция, характеризующая очаг возникновения болезни $F_{tum}$, $\alpha$ -- параметр сопротивления организма развитию опухоли.
%Величина $\alpha$ может задавать зависимость системы от применения какой-либо терапии, или естественной сопротивляемости болезни имунной системы, или клиренса.

%\subsection{Данные клинических исследований}
\subsection{Clinical study data}

%Мы использовали данные, опубликованные в \cite{Gui}, для валидации наших математической и численной моделей. 
We used the data published in the article \cite{Gui} to validate our mathematical and numerical models.
%Представленные данные содержат информацию об изменении размеров раковых опухолей для 10 пациентов.
The presented data contain information about change in  size of cancerous tumors for 10 patients.
%Измерения проводились на основе серийных исследований магнитно-резонансной томографии (MRI). 
The measurements were carried out on basis of serial magnetic resonance imaging (MRI) studies. 
%Объемы опухоли определялись путем ручной сегментации на серийных изображениях для изучения естественного течения и роста области поражения.
Tumor volumes were determined by manual segmentation on serial image to study natural course and growth of lesion.
%Учитывая особенности использованного метода, полученные данные обладают погрешностью, обусловленной, как применяемым методом медицинского обследования, так и алгоритмами обработки изображений. 
The data obtained have an error due to both applied method of medical examination and image processing algorithms.
%Тем не менее полученные данные на качественном уровне хорошо описывают процесс роста объема опухолевой ткани. 
Nevertheless, data obtained qualitatively describe well the process of tumor tissue volume growth.

%\subsection{Численные методы и вычислительные технологии}
\subsection{Numerical methods and computational technologies}
%Для численного решения уравнения \ref{eq:equation2.1} используется трехмерная многослойная схема. 
A three-dimensional multilayer scheme is used for numerical solution of the equation Eq. \ref{eq:equation2.1}.
%Мы рассматриваем задачу в декартовой системе координат $(x, y, z)$. 
We consider problem in the Cartesian coordinate system $(x, y, z)$.
%Рассматриваемая область интегрирования покрывается фиксированной в пространстве (эйлеровой) расчетной трехмерной сеткой, которая представляет собой набор прямоугольных ячеек со сторонами равными длине $\Delta x$, $\Delta y$ и $\Delta z$. 
The integration region under consideration is covered by a spatially fixed (Eulerian) computational three-dimensional grid, which is a set of rectangular cells with sides equal to the length $\Delta x$, $\Delta y$ and $\Delta z$.
%Расчетная сетка является равномерной, то есть размер ячеек постоянен во всей области задачи. 
Computational grid is uniform, that is, cell size is constant throughout area of problem.
%Центр ячейки соответственно будет находится в точке с координатами $i$, $j$ и $k$, которые являются целыми числами. 
Center of cell, respectively, will be located at point with coordinates $i$, $j$ and $k$, which are integers.
%Переменные $\phi = (D, u, v, w, Q, P, K)$ являются дискретными и ассоциируются с узлами пространственной сетки, которая задана в виде $x_i, y_j, z_k$, и с узлами временной сетки, которая представляется в виде $t_n: \phi(x_i, y_j, z_k, t_n)\equiv \phi^{n}_{i,j,k}$. 
Variables $\phi = (D, u, v, w, Q, P, K)$ are discrete and are associated with nodes of spatial grid, which is given in form $x_i, y_j, z_k$ and with nodes of time grid, which is represented in  form $t_n: \phi(x_i, y_j, z_k, t_n)\equiv \phi^{n}_{i,j,k}$.
The time grid step varies, i.e. it is non-uniform ($t_{n+1} = t_{n} + \Delta t_{n}$), in contrast, as already noted, step in space does not change from time step ($x_{i+1} = x_{i} + \Delta x, y_{j+1} = y_{j} + \Delta y, z_{k+1} = z_{k} + \Delta z$). 
%В процессе аппроксимации решаемого уравнения (\ref{eq:equation2.1}) получаем явную разностную схему для его решения относительно искомого значения $P$ в декартовой системе координат ($x, y, z$) для каждой ячейки в ней ($i, j, k$):
In the process of approximating Eq. (\ref{eq:equation2.1}) being solved, we obtain an explicit difference scheme for solving it with respect to desired value $P$ in the Cartesian coordinate system ($x, y, z$) for each cell in it ($i , j, k$):
\begin{eqnarray}
    \label{eq:equation2.11}
    D^{n+1}_{i,j,k} = D^{n}_{i,j,k} - \frac{1}{\Delta x \Delta y \Delta z} (\Delta M^{n}_{i+1/2,j,k} - \Delta M^{n}_{i-1/2,j,k} +\nonumber\\ 
    + \Delta M^{n}_{i,j+1/2,k} - \Delta M^{n}_{i,j-1/2,k} + \Delta M^{n}_{i,j,k+1/2} - \Delta M^{n}_{i,j,k-1/2})+\nonumber\\ 
    + \frac{\Delta t_{n}}{(\Delta x)^{2}} \left(P_{i+1/2,j,k} \left(D^{n}_{i+1,j,k} - D^{n}_{i,j,k}\right) - P_{i-1/2,j,k} \left(D^{n}_{i,j,k} - D^{n}_{i-1,j,k}\right)\right) + \nonumber\\ 
    + \frac{\Delta t_{n}}{(\Delta y)^{2}} \left(P_{i,j+1/2,k} \left(D^{n}_{i,j+1,k} - D^{n}_{i,j,k}\right) - P_{i,j-1/2,k} \left(D^{n}_{i,j,k} - D^{n}_{i,j-1,k}\right)\right) + \nonumber\\ 
    + \frac{\Delta t_{n}}{(\Delta z)^{2}} \left(P_{i,j,k+1/2} \left(D^{n}_{i,j,k+1} - D^{n}_{i,j,k}\right) - P_{i,j,k-1/2} \left(D^{n}_{i,j,k} - D^{n}_{i,j,k-1}\right)\right) + \nonumber\\ 
    + \Delta t_{n}(Q^{n}_{i,j} - \alpha D^{n}_{i,j,k}).
\end{eqnarray}
%Переменные с дробными индексами здесь относятся к границам расчетных ячеек, и определяются следующим образом:
Variables with fractional indices here refer to boundaries of calculated cells, and are defined as follows:
\begin{eqnarray}
    \label{eq:equation2.12}
    P^{n}_{i+1/2,j,k} = P(| \overrightarrow{S}|^{n}_{i+1/2, j, k}), | \overrightarrow{S}|^{n}_{i+1/2, j, k} = \frac{| \overrightarrow{S}|^{n}_{i, j, k} + | \overrightarrow{S}|^{n}_{i\pm 1, j, k}}{2},
    \end{eqnarray}
    
\begin{eqnarray}
        P^{n}_{i,j+1/2,k} = P(| \overrightarrow{S}|^{n}_{i, j+1/2, k}), | \overrightarrow{S}|^{n}_{i, j+1/2, k} = \frac{| \overrightarrow{S}|^{n}_{i, j, k} + | \overrightarrow{S}|^{n}_{i, j\pm 1, k}}{2},
        %\nonumber\\
\end{eqnarray}

\begin{eqnarray}
    P^{n}_{i,j,k+1/2} = P(| \overrightarrow{S}|^{n}_{i, j, k+1/2}), | \overrightarrow{S}|^{n}_{i, j, k+1/2} = \frac{| \overrightarrow{S}|^{n}_{i, j, k} + | \overrightarrow{S}|^{n}_{i, j, k\pm 1}}{2}.
\end{eqnarray}
%Следует задавать параметры, в частности значения величины $P$ корректными для данной решаемой задачи, для достаточного описания дианмической системы на рассматриваемом уровне детализации и приближения. 
It is necessary to set the parameters, in particular, the values of $P$ to be correct for given problem being solved, for a sufficient description of  dynamic system at considered level of detail and approximation.
%Приведем конечно-разностную аппроксимацию для значений коэффициентов для нашей задачи
We present a finite-difference approximation for the values of the coefficients for our problem
\begin{eqnarray}
    \label{eq:equation2.13}
    \Delta M^{n}_{i+1/2,j,k} = \Delta y \Delta z \Delta t_n 
        \begin{cases}
            D^{n}_{i,j,k}u^{n}_{i+1/2,j,k}, u^{n}_{i+1/2,j,k} > 0\\
            D^{n}_{i+1,j,k}u^{n}_{i+1/2,j,k}, u^{n}_{i+1/2,j,k} < 0
        \end{cases}, 
\end{eqnarray}
\begin{eqnarray}
    \Delta M^{n}_{i-1/2,j,k} = \Delta y \Delta z \Delta t_n 
        \begin{cases}
            D^{n}_{i,j,k}u^{n}_{i-1/2,j,k}, u^{n}_{i-1/2,j,k} > 0\\
            D^{n}_{i-1,j,k}u^{n}_{i-1/2,j,k}, u^{n}_{i-1/2,j,k} < 0
        \end{cases}, 
\end{eqnarray}
\begin{eqnarray}
    \Delta M^{n}_{i,j+1/2,k} = \Delta x \Delta z \Delta t_n 
        \begin{cases}
            D^{n}_{i,j,k}v^{n}_{i,j+1/2,k}, v^{n}_{i,j+1/2,k} > 0\\
            D^{n}_{i,j+1,k}v^{n}_{i,j+1/2,k}, v^{n}_{i,j+1/2,k} < 0
        \end{cases}, %\tag{\stepcounter{equation}\theequation}
\end{eqnarray}
\begin{eqnarray}
    \Delta M^{n}_{i,j-1/2,k} = \Delta x \Delta z \Delta t_n 
        \begin{cases}
            D^{n}_{i,j,k}v^{n}_{i,j-1/2,k}, v^{n}_{i,j-1/2,k} > 0\\
            D^{n}_{i,j-1,k}v^{n}_{i,j-1/2,k}, v^{n}_{i,j-1/2,k} < 0
        \end{cases},
\end{eqnarray}
\begin{eqnarray}
    \Delta M^{n}_{i,j,k+1/2} = \Delta x \Delta y \Delta t_n 
        \begin{cases}
            D^{n}_{i,j,k}w^{n}_{i,j,k+1/2}, w^{n}_{i,j,k+1/2} > 0\\
            D^{n}_{i,j,k+1}w^{n}_{i,j,k+1/2}, w^{n}_{i,j,k+1/2} < 0
        \end{cases}, 
\end{eqnarray}
\begin{eqnarray}
    \Delta M^{n}_{i,j,k-1/2} = \Delta x \Delta y \Delta t_n 
        \begin{cases}
            D^{n}_{i,j,k}w^{n}_{i,j,k-1/2}, w^{n}_{i,j,k-1/2} > 0\\
            D^{n}_{i,j,k-1}w^{n}_{i,j,k-1/2}, w^{n}_{i,j,k-1/2} < 0
        \end{cases} ,
\end{eqnarray}
where $u^{n}_{i\pm1/2,j,k}=\frac{u^{n}_{i,j,k} + u^{n}_{i\pm1,j,k}}{2}$, $v^{n}_{i,j\pm1/2,k}=\frac{v^{n}_{i,j,k} + v^{n}_{i,j\pm1,k}}{2}$ and $w^{n}_{i,j,k\pm1/2}=\frac{w^{n}_{i,j,k} + w^{n}_{i,j,k\pm1}}{2}$. 
%Потоки через границы построенных эйлеровых ячеек, обусловленные силовым переносом, выражаются в данной схеме через значение $\Delta M^{n}$.
Flows through boundaries of the constructed Euler cells, due to force transfer, are expressed in this scheme through the value $\Delta M^{n}$.

%Как и любая явная схема, численная схема (\ref{eq:equation2.11}) устойчива лишь условно. Анализ устойчивости схемы (\ref{eq:equation2.13}) дает точную границу ее применения:
This numerical scheme (\ref{eq:equation2.11}) is only conditionally stable, like any explicit scheme.
%Анализ устойчивости схемы (\ref{eq:equation2.13}) дает точную границу ее применения:
Analysis of stability of scheme (\ref{eq:equation2.13}) gives exact limit of its application:
\begin{equation}
    \Delta t_{n} = K min\left(\frac{(\Delta x)^2}{2P^{n}_{i,j,k}} + \frac{\Delta x}{u^{n}_{i,j,k}},\frac{(\Delta y)^2}{2P^{n}_{i,j,k}} + \frac{\Delta y}{y^{n}_{i,j,k}},\frac{(\Delta z)^2}{2P^{n}_{i,j,k}} + \frac{\Delta z}{z^{n}_{i,j,k}}\right),
\end{equation}
where $K$ is Courant number. 
%Во всех приведенных в статье случаях вычисления проводились при значениях числа Куранта $K=0.1$, которое было подобрано на основе проведенных экспериментов.
In all the cases presented in this article, the calculations were carried out for the values of the Courant number $K=0.1$, which was selected on basis of experiments performed.

\begin{figure}[h!] 
  \center
  \includegraphics [width = \linewidth] {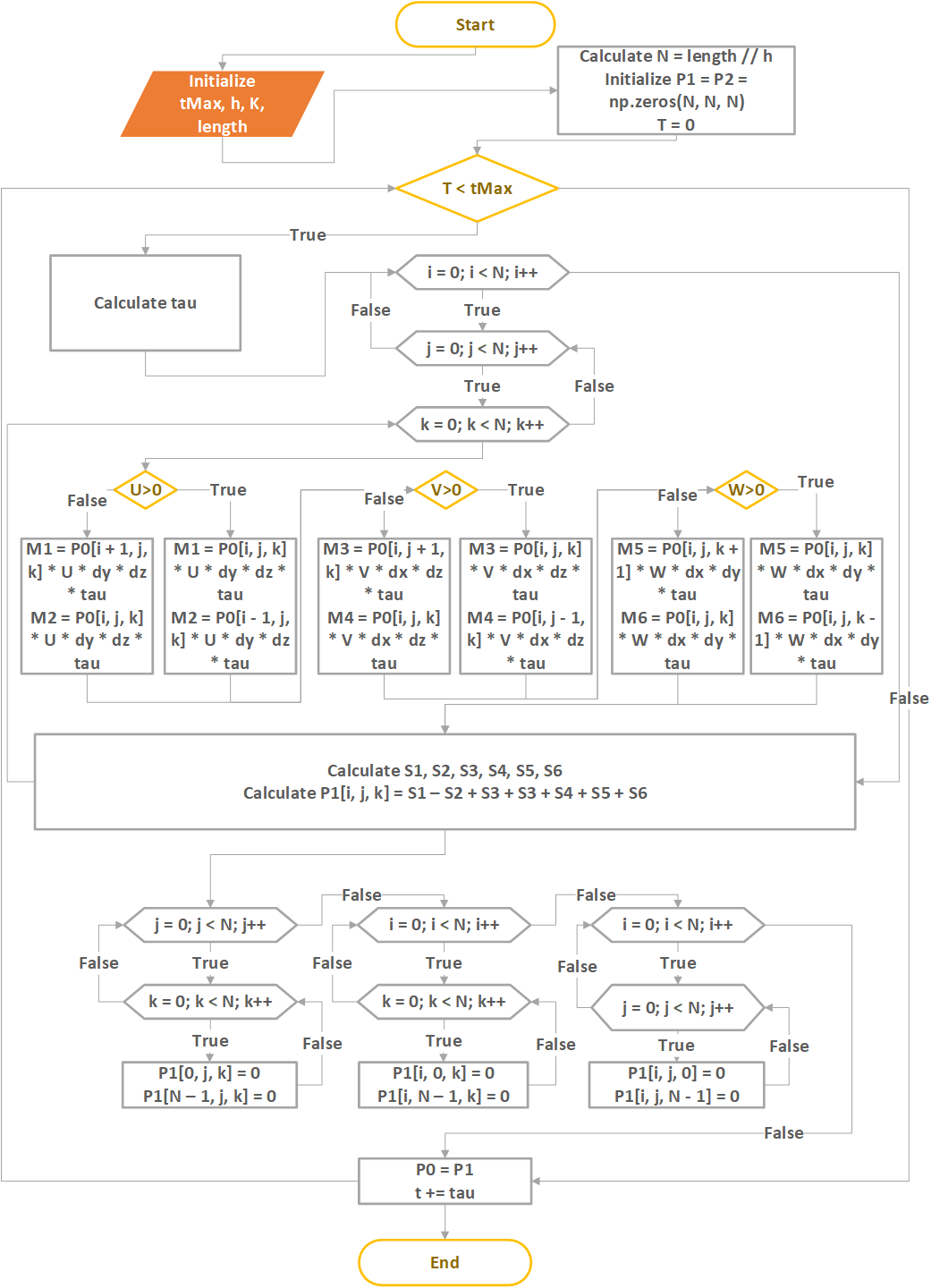}
 % \caption{Блок-схема реализации численного метода, используемого для интегрирования уравнения диффузии} 
 \caption{The block diagram of implementation of numerical method used to integrate the diffusion equation.}
  \label{img:block_schemeENG}  
\end{figure}

%Для численного решения уравнения \ref{eq:equation2.1} на основе приведенной конечно-разностной аппроксимации было разработано программное обеспечение на языках программирования C \# и Python. 
Software was developed in programming languages C\# and Python for numerical solution of the Eq. \ref{eq:equation2.1} based on given finite difference approximation.
%Блок-схема разработанного программного комплекса представлена на рисунке \ref{img:block_schemeENG}. 
The block diagram of developed software package is shown in Fig. \ref{img:block_schemeENG}.

%После инициализации входных параметров модели вычисляется количество расчетных ячеек $N$ на основе шага по пространству $h$. 
Number of cells $N$ is calculated after initializing input parameters of model based on the spatial step $h$.
%Далее запускается цикл по времени, который выполняется до тех пор, пока время не достигнет заданного значения $t_{max}$. 
Next, a time loop is started, which is executed until time reaches the specified value $t_{max}$.
%В цикле вычисляются промежуточные величины конечно-разностной схемы. 
In cycle intermediate values of  finite difference scheme are calculated.
%С использованием этих значений на выходе мы получаем концентрацию раковых клеток в некоторой области.
At output we get the concentration of cancer cells in some area using these values.

%\section{Результаты вычислительных экспериментов по моделированию роста опухоли} \label{NumSim}
\section{Results of computational experiments on simulation tumor growth} \label{NumSim}
%\subsection{Поиск начальных параметров математической модели на основе данных клинических исследований}
\label{subsec_31}
\subsection{Search for initial parameters of a mathematical model based on clinical observation data}
\label{subsec_31}
%В этом разделе мы определим границы варьируемых параметров, используемых для исследования роста опухоли. 
We define the boundaries of the variable parameters used to study tumor growth in this section.
%Найдем параметр $S$, определяющий скорость распространения заболевания. 
We find the parameter $S$, which determines rate of spread of disease.
%Этот параметр может зависеть от многих факторов, таких как возраст пациента, пол, генетическая предрасположенность и т.д. 
This parameter may depend on many factors, such as the patient's age, gender, genetic predisposition, etc.
%Результаты моделирования мы будем сравнивать с данными пациентов, у которых была обнаружена глиома -- первичная опухоль головного мозга \cite{Louis}. 
We will compare simulation results with data from patients who have been diagnosed with glioma, a primary brain tumor \cite{Louis}.
%Она имеет множество подтипов, которые на данном этапе в математической модели не учитываются. 
It has many subtypes, which are not taken into account in the mathematical model at this stage.
%Для семейства таких опухолей, кроме прочего, к критериям риска следует отнести еще и подтип астроцитомы и наличие неврологического дефицита \cite{Pignatti}. 
The risk criteria for a family of such tumors should also include, among other things, astrocytoma subtype and presence of a neurological deficit \cite{Pignatti}. 
%В работе \cite{Gui} по статистическому критерию Манна—Уитни было выявлено, что у людей моложе 50 лет развитие болезни протекает гораздо медленнее \cite{Korneev}. 
In the article\cite{Gui}, according to the Mann-Whitney statistical test, it was found that in people under 50 years of age, development of disease proceeds much more slowly \cite{Korneev}.
%У пациентов женского пола наблюдались также тенденции к более низким степеням роста в сравнении с пациентами мужского пола. 
Trends towards lower growth rates compared to male patients were also observed in female patients.

%Анализ изменения объема опухоли является более информативным, чем анализ изменения ее диаметра. 
Analysis of changes in tumor volume is more informative than analysis of changes in its diameter.
%Это вытекает из того факта, что опухоль в момент ее выявления и с течением времени имеет не строго геометрическую, хотя и в общим чертах эллипсоидную и продолговатую форму. 
This follows from the fact that tumor, at the time of its discovery and over time, is not strictly geometric, although in general terms it is ellipsoid and oblong.
%Анализ диаметра не дает нам той точности измерений, что анализ объема, так как он еще и не учитывает достаточно сильной неровности границ опухоли (see Fig. \ref{img:heterogeneoustumor}). 
Diameter analysis does not give us the accuracy of measurements that volume analysis does, since it also does not take into account sufficiently strong roughness of tumor boundaries (see Fig. \ref{img:heterogeneoustumor}).
%Стоит отметить, что исследование болезни при варьировании общего направления развития опухоли рассматривать не будем, так как болезнь может развиваться в тех направлениях, предсказать которые весьма затруднительно.
It should be noted that we will not consider study of disease with variation in the general direction of tumor development, since disease can develop in those directions that are very difficult to predict.
%То есть в нашем приближении направление задается случайным образом.
That is, direction is given randomly in our approximation.
%Кроме того, скорость распространения заболевания в различных тканях организма также может значительно изменяться. 
In addition, rate of spread of disease in different tissues of  body can also vary significantly.
%Также мы определяем коэффициент диффузии $P$, который в нашей модели является свободным параметром. 
We also determine the diffusion coefficient $P$, which is a free parameter in our model.

%Для нахождения параметров $P$ и $S$ мы применяем метод наименьших квадратов. 
We apply the least squares method to find the parameters $P$ and $S$.
%Этот метод основывается на минимизации суммы квадратов отклонений результатов моделирования от результатов клинического обследования пациента в каждый момент времени. 
This method is based on minimizing sum of squared deviations of simulation results from the results of clinical examination of patient at each point in time.
%В результате для 10 пациентов были получены значения начальных параметров моделей
As a result, the values of initial parameters of models were obtained for 10 patients (Table \ref{table:DataS}).

\begin{table}[h!]
\begin{center}
   % \caption{\label{table:DataS} Значения коэффициента скорости распространения заболевания $S$ и коэффициента диффузии $P$ для каждого вычислительного эксперимента, соответствующего отдельно взятому пациенту}
    \caption{\label{table:DataS} The values of disease spread rate coefficient $S$ and the diffusion coefficient $P$ for each computational experiment corresponding to a single patient.}
\begin{tabular}{lcc}
\hline
            & \multicolumn{1}{l}{$P$ (mkm$^2$/s)} & \multicolumn{1}{l}{$S$ (mkm/s)} \\ \hline
Model 1 (a) & 2.2                           & 1.9                           \\
Model 2 (b) & 2.45                          & 2.36                          \\
Model 3 (c) & 1.88                          & 2.34                          \\
Model 4 (d) & 3.73                          & 3.14                          \\
Model 5 (e) & 3.51                          & 2.88                          \\
Model 6 (f) & 3.67                          & 3.22                          \\
Model 7     & 1.81                          & 1.8                           \\
Model 8     & 2.51                          & 1.69                          \\
Model 9     & 3.89                          & 2.79                          \\
Model 10    & 3.47                          & 2.96                          \\ \hline
\end{tabular}
\end{center}
\end{table}

\begin{figure}[h!] 
 \center
  \includegraphics [width = 0.6\linewidth] {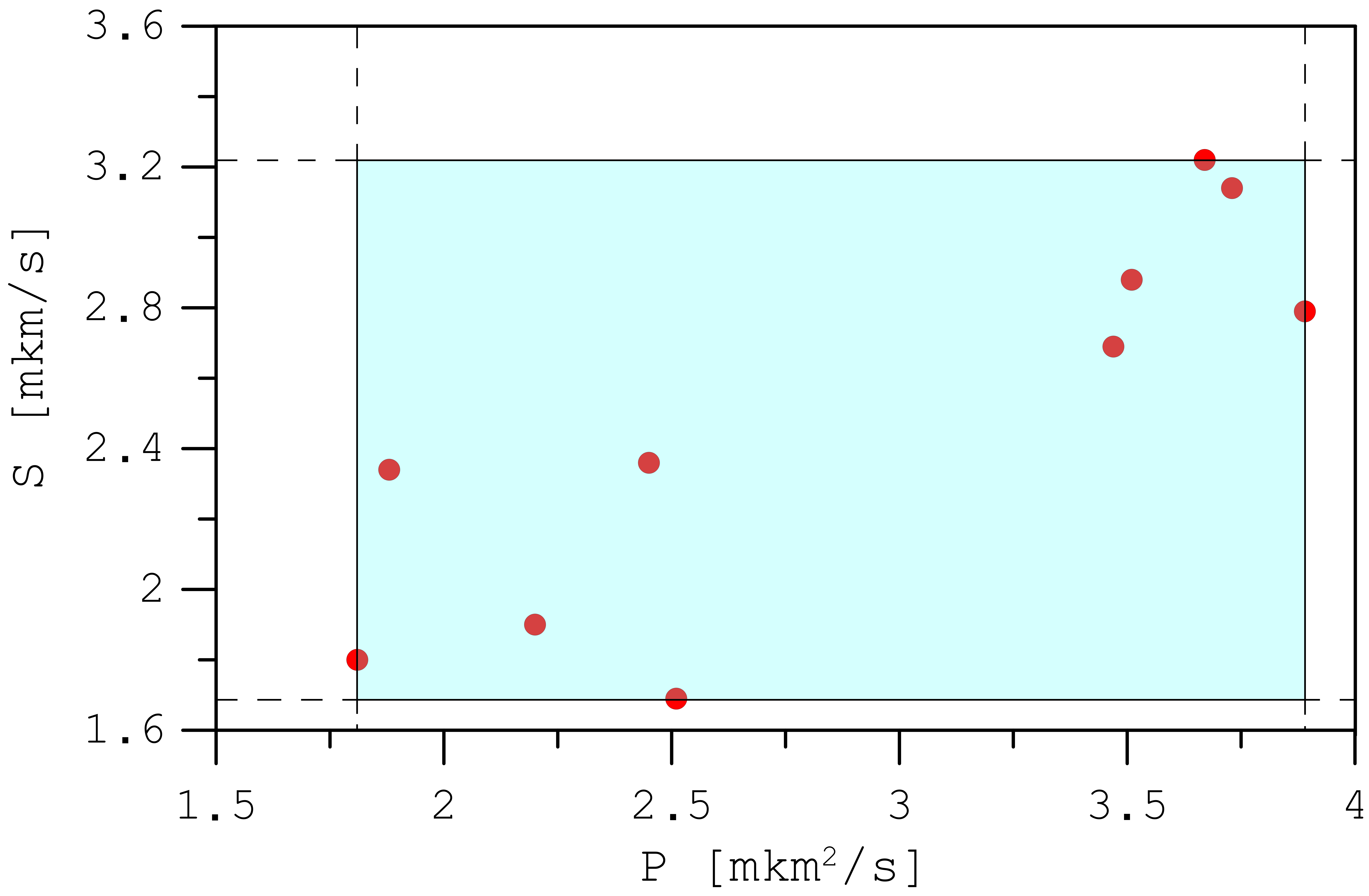}
   %\caption{Область допустимых значений свободных параметров $P$ и $S$ (показана голубым цветом). Красными точками показаны значения параметров для 10 моделей, соответствующих реальным пациентам.} 
   \caption{The range of admissible values of the free parameters $P$ and $S$ (shown in blue). Red dots show parameter values for 10 models corresponding to real patients.} 
  \label{img:parameterplane}  
\end{figure}

%Отметим, в каких границах могут находиться значения свободных параметров модели. 
Note the limits within which values of free parameters of  model can lie.
%На рисунке \ref{img:parameterplane} голубым цветом показана область, в которой могут лежать значения величин $P$ и $S$ для 10 рассматриваемых пациентов. 
The blue color in the Fig. \ref{img:parameterplane} shows area in which the values of $P$ and $S$ for the 10 considered patients may lie.
%В результате мы получаем, что диапазон значений для модуля скорости $S$: 1.69 mkm/s $\leq$ $S$ $\leq$ 3.22 mkm/s, 
As a result, we get that range of values for the modulus of speed $S$: 1.69 mkm/s $\leq$ $S$ $\leq$ 3.22 mkm/s, 
%а для значений коэффициента распространения опухоли $P$: 1.81 mkm$^2$/s $\leq$ $P$ $\leq$ 3.89 mkm$^2$/s. 
and for the values of tumor spread coefficient $P$: 1.81 mkm$^2$/s $\leq$ $P$ $\leq$ 3.89 mkm$^2$/s. 

%Средние выборочные значения параметров для <<типичного>> пациента равны $P \approx$ 2.9 mkm$^2$/s, а скорости распространения болезни $S \approx$ 2.5 mkm/s. 
The mean sample values of the parameters for a ''typical'' patient are equal to $P \approx$ 2.9 mkm$^2$/s, and the rate of spread of disease $S \approx$ 2.5 mkm/s.
%Зная границы значений этих параметров, можно применять данную модель для исследования динамики рака.
This model can be used to study the dynamics of cancer, knowing limits of the values of these parameters.

%\subsection{Исследование динамики объема опухоли}
\subsection{Study of tumor volume dynamics}
\begin{figure}[h!] 
  \center
  \includegraphics [width = \linewidth] {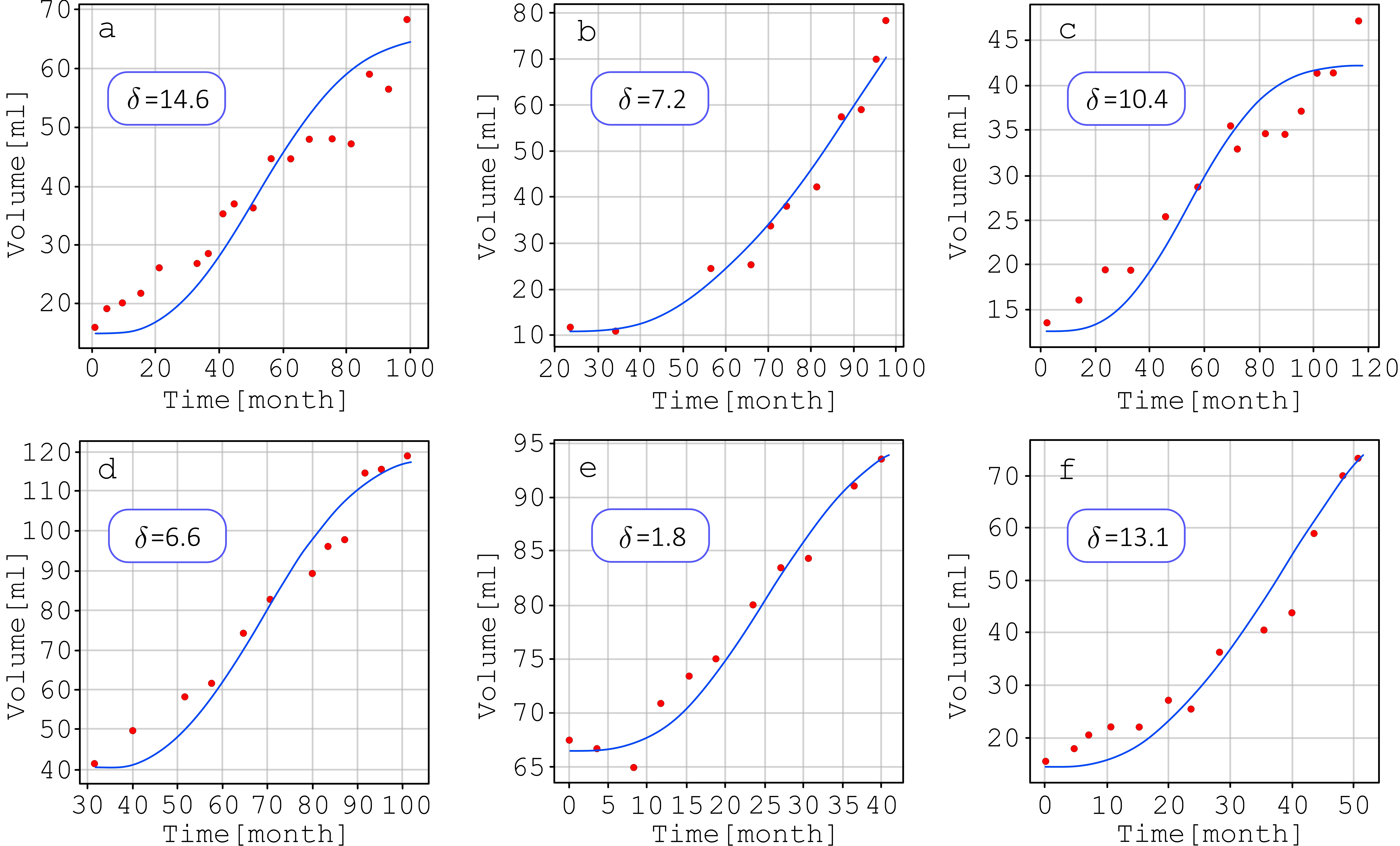}
  % \caption{Результаты моделирования динамики роста раковой опухоли для 6 пациентов из выборки данных при различном наборе входных параметров (см. таблицу \ref{table:DataS}) с данными реальных медицинских обследований \cite {Gui}. Во всех случаях относительная погрешность $\delta$ указана в процентах.} 
    \caption{The results of modeling growth dynamics of a cancerous tumor for 6 patients from a data sample with a different set of input parameters (see. Table \ref{table:DataS}) with data from real medical examinations \cite {Gui}. In all cases, the relative error $\delta$ is given as a percent.}
  \label{img:theBestResultsAnyVolume1}  
\end{figure}

%Для исследования динамики объема опухоли в вычислительных экспериментах были заданы временные промежутки, соответствующие периоду развития болезни в клинических данных. 
Time intervals corresponding to the period of disease development in clinical data were set in computational experiments to study dynamics of tumor volume.
%Каждому зафиксированному состоянию пациента соответствует точки красного цвета (рисунок \ref{img:theBestResultsAnyVolume1}). 
Red dots correspond to each recorded patient condition (Fig. \ref{img:theBestResultsAnyVolume1}). 
%Наблюдения изменения объема опухоли во всех случаях велись до начала лечения. 
Observations of changes in tumor volume in all cases were carried out before the start of treatment.
%Отметим, что данные о дальнейшей динамике объема опухоли отсутствуют.
Note that there are no data on further dynamics of tumor volume.
%Для каждого вычислительного эксперимента была рассчитана относительная погрешность на основе приведенных клинических случаев
The relative error was calculated for each computational experiment based on the given clinical cases
\begin{eqnarray}
    \label{error}
    \delta = \sum\limits_{i=1}^n \frac{V^{i}_{sim}-V^{i}_{obs}}{V^{i}_{obs}}\cdot 100 \%,
\end{eqnarray}
where $V^{i}_{sim}$ is simulated value of tumor volume in $i$ moment of time, $V_{obs}$ is the value of tumor volume for data of clinical observations in $i$ moment of time, $n$ is number of observations.
\begin{figure}[h!] 
  \begin{center}
  \includegraphics [width = 0.6\linewidth] {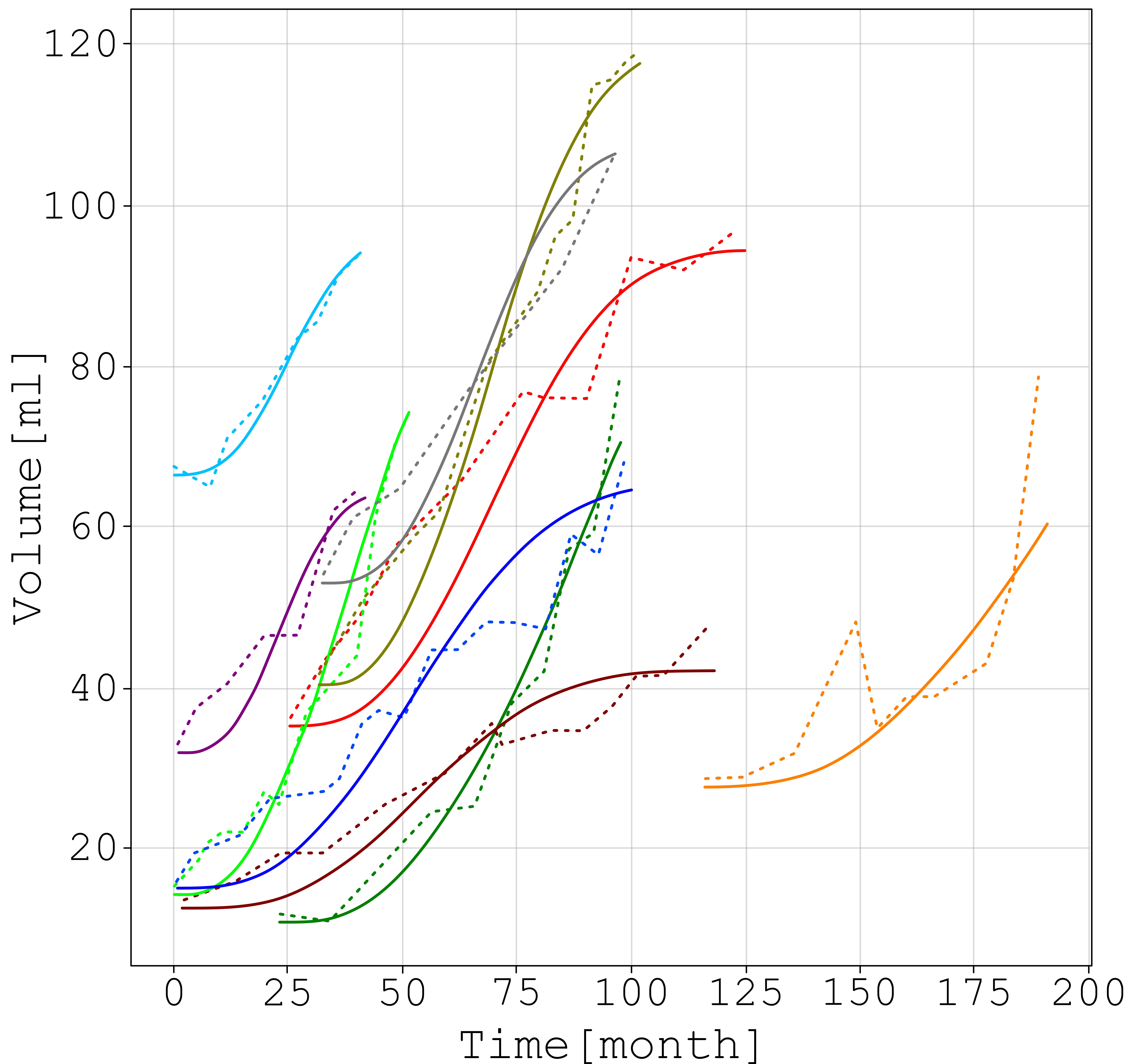}
  %\caption{Смоделированная динамика (показаны цветными сплошными линиями) изменения объема опухоли, соответствующая данным о 10 реальных пациентах (показаны цветными пунктирными линиями)} 
  \caption{The simulated dynamics (shown in colored solid lines) of changes in tumor volume corresponding to data on 10 real patients (shown in colored dotted lines)}
  \label{img:AllPatients}
  \end{center}
\end{figure}
%\begin{table}[h!]
%    \begin{center}
%    \caption{\label{table:DataRelativeError}Величина погрешности моделирования для каждого клинического случая %\textcolor{red}{!!!! убрать и перенести на рисунки}} 
%    \begin{tabular}{ | l | l | l | l | }
 %   \hline
 %   № Пациента & Погрешность & № Пациента & Погрешность\\ \hline
%    1 & 14.61\% & 6 & 10.40\%\\
%    2 & 9.77\% & 7 & 13.03\%\\
%    3 & 7.28\% & 8 & 6.87\%\\
%    4 & 10.78\% & 9 & 6.58\%\\
%    5 & 8.60\% & 10 & 1.86\%\\
%    \hline
%    \end{tabular}
%    
%    \end{center}
%\end{table}

%На основе построенных решений мы наблюдаем, что для случаев a and f (см. рисунок \ref{img:theBestResultsAnyVolume1}) характерно самое большое расхождение между смоделированными данными и данными реальных клинических исследований. 
Based on constructed solutions, we observe that cases a and f are characterized by largest discrepancy between simulated data and data from real clinical trials.
%В указанных случаях присутствует резкое малопредсказуемое изменение тенденции к росту опухоли, проявляющееся в разные моменты времени.
A sharp unpredictable change in the trend towards tumor growth, which manifests itself at different points in time, is present in these cases.
%При этом наибольшее отклонение наблюдается для случая a. 
The largest deviation is observed for the case a.
%Этот пациент характеризуется самым медленным распространением болезни, которое меняет свою скорость примерно на 70-ом месяце с начала наблюдений.
This patient is characterized by the slowest spread of the disease, which changes its speed at about the 70th month from the start of observations.
%Однако и здесь отклонение в начале наблюдений практически отсутсвует, а в конце наблюдений оно не превышает $70$ мл. 
However, even here deviation at the beginning of observations is practically absent, and at the end of observations it does not exceed $70$ ml.
%Наиболее точные результаты наблюдаются у случая e.
The most accurate results are observed in the case e.
%Для наглядности приведем также общую сводку с динамикой болезни у пациентов (рис. \ref{img:AllPatients}) по сравнению с клиническими данными.
For clarity, we also give a general summary of dynamics of  disease in patients (see Fig. \ref{img:AllPatients}) compared with clinical data.
%\linewidth

%\begin{table}[h!]
%    \begin{center}
%    \caption{\label{table:DataS}Значения коэффициента распространения болезни $S$ для каждого пациента}
%    \begin{tabular}{ | l | l | l | l | }
%    \hline
%    № Пациента & Значение $S$ & № Пациента & Значение $S$ \\ \hline
%    1 & 1.9  & 6 & 2.34\\
%    2 & 1.8  & 7 & 3.22\\
%    3 & 2.36  & 8 & 2.96\\
%    4 & 1.69  & 9 & 3.14\\
%    5 & 2.79  & 10 & 2.88\\
%    \hline
%    \end{tabular}
     
%    \end{center}
%\end{table}

%\begin{table}[h!]
%    \begin{center}
%    \caption{\label{table:DataD}Значения коэффициента $S$ для каждого пациента}
%    \begin{tabular}{ | l | l | l | l | }
%    \hline
%    № Пациента & Значение скорости $D$ & № Пациента & Значение скорости $D$ \\ \hline
%    1 & 2.2  & 6 & 1.88\\
%    2 & 1.81  & 7 & 3.67\\
%    3 & 2.45  & 8 & 3.47\\
%    4 & 2.51  & 9 & 3.73\\
%    5 & 3.89  & 10 & 3.51\\
%    \hline
%    \end{tabular}
     
%    \end{center}
%\end{table}

\subsection{Study of spatial structure of tumor}
\begin{figure}[h!] 
  \center
  \includegraphics [width = 1 \linewidth] {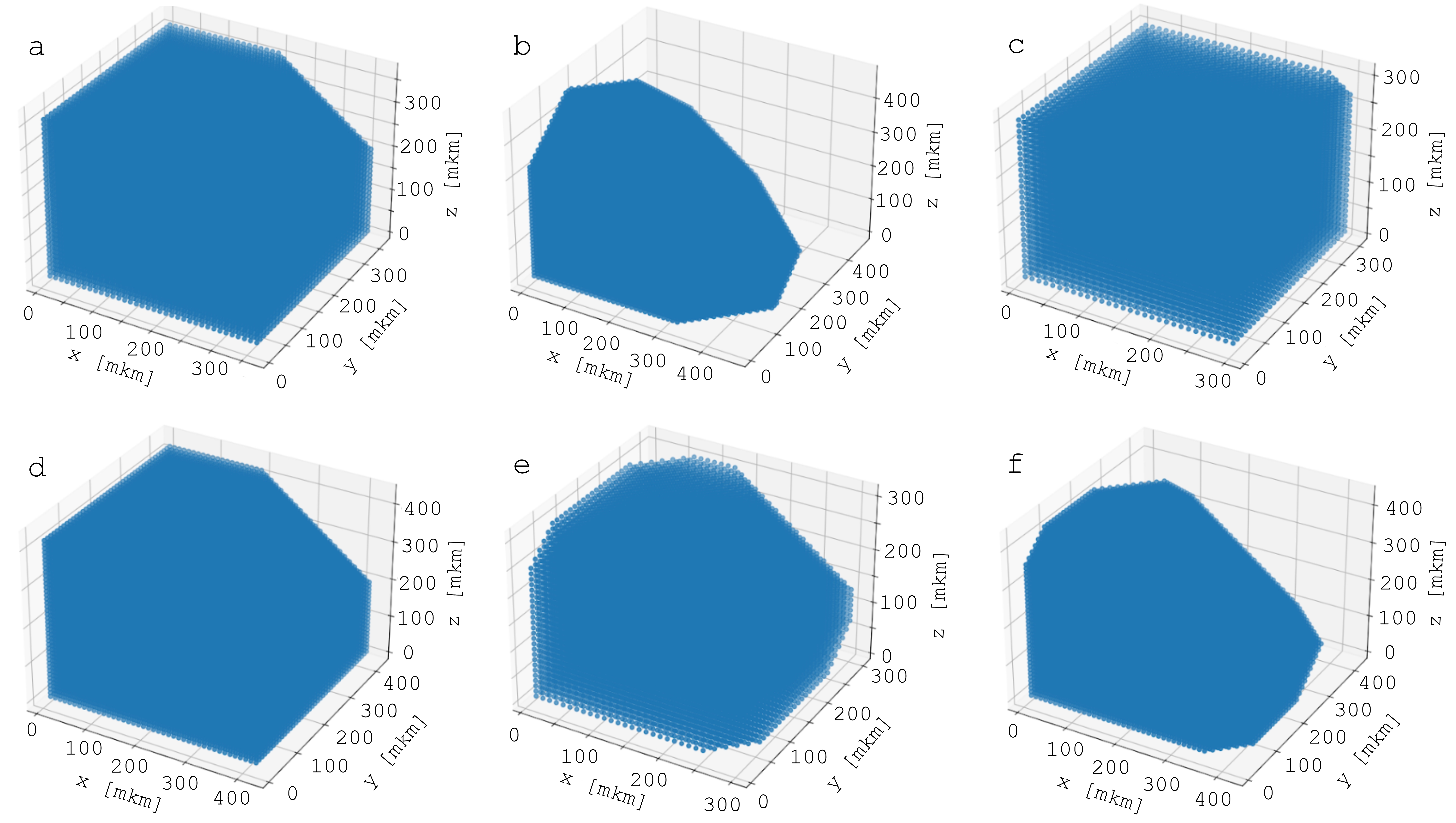}
   %\caption{Пространственная трехмерная структура опухоли для 6 пациентов из выборки данных при различном наборе входных параметров (см. таблицу \ref{table:DataS}) к конечный момент расчетного времени}
   \caption{The spatial three-dimensional structure of the tumor for 6 patients from data sample with a different set of input parameters (see. Table \ref{table:DataS}) at the end point of the calculated time.}
  \label{3D}  
\end{figure}

%При вычислении объема пораженной опухолью ткани, вообще говоря, надо учитывать форму этой области. 
Generally speaking, shape of this area must be taken into account when calculating volume of tissue affected by the tumor.
%Мы построили трехмерную структуру опухоли для всех наблюдаемых случаев с учетом фактора неоднородности среды (рисунок \ref{3D}). 
We constructed a three-dimensional structure of the tumor for all observed cases, taking into account environmental heterogeneity factor (Fig. \ref{3D}).
%Видно, что размеры и форма области, пораженной раком, отличается для всех моделей, в том числе за счет различной структуры биологических тканей.
It can be seen that size and shape of area affected by cancer differs for all models, including due to the different structure of biological tissues.
%Так как период деления раковых клеток короткий, то раковые клетки в процессе эволюции системы быстро наращивают занимаемый ими объем. 
Cancer cells in process of evolution of the system rapidly increase volume they occupy, since period of division of cancer cells is short.
%Оценка структуры опухоли является важной, например, для определения эффективности хирургического лечения пациентов.
Evaluation of structure of tumor is important, for example, to determine effectiveness of surgical treatment of patients.

%\section{Обсуждение и заключение} \label{concldiss}
\section{Discussion and conclusion} \label{concldiss}

%Медицинская практика основана на разнообразных методах диагностики злокачественных опухолей (ultrasound, mammography, tomography).
Medical practice is based on a variety of methods for diagnosing malignant tumors (ultrasound, mammography, tomography).
%Несмотря на развитие этих методов, ни один из них не гарантирует раннее выявление рака. 
Despite the development of these methods, none of them guarantees early detection of cancer.
%Совместное применение диагностических техник и методов математического моделирования может помочь в решении этой проблемы. 
The combined use of diagnostic techniques and mathematical modeling methods can help in solving this problem.
%Наши основные усилия направлены на создание универсальных методов компьютерного моделирования биофизических процессов. 
Our main efforts are aimed at creating universal methods for computer simulation of biophysical processes.
%Такие методы могут повысить качество медицинской диагностики с использованием алгоритмов машинного обучения. 
Such methods can improve the quality of medical diagnostics using machine learning algorithms.
%На сегодняшний день активно развивается метод диагностики онкологических заболеваний по данным микроволновой радиотермометрии \cite{PolyakovPopov,LiGalazis}. 
To date, a method for diagnosing oncological diseases based on microwave radiothermometry data is being actively developed \cite{PolyakovPopov,LiGalazis}.
%Применение этого метода основано на анализе тепловых полей отдельных органов пациента. 
The application of this method is based on the analysis of thermal fields of individual organs of the patient.
%В основе диагностики рака лежит наблюдение того, что наиболее опасные и быстрорастущие опухоли характеризуются высоким удельным тепловыделением \cite{Gautherie-1982}. 
The diagnosis of cancer is based on the observation that the most dangerous and rapidly growing tumors are characterized by high specific heat \cite{Gautherie-1982}.
%Таким образом опухоли оказывают влияние на тепловые поля в биологических тканях. 
Thus, tumors affect thermal fields in biological tissues.
%Поэтому учет удельного тепловыделения опухоли в математической модели видится актуальным направлением развития нашего исследования.
Therefore, taking into account the specific heat release of the tumor in the mathematical model seems to be an important direction in the development of our study.

%Функция $Q$ в уравнении (\ref{eq:equation2.1}), которая отвечает за появление опухоли, нуждается в дальнейшей математической проработке.
The function $Q$ in the Eq. (\ref{eq:equation2.1}), which is responsible for the appearance of a tumor, needs further mathematical elaboration.
%В этой работе значение $Q$ задается равной константе, и не зависит от каких-либо других переменных. 
In this article, the value of $Q$ is set to a constant, and does not depend on any other variables.
%Исследование процесса появления опухоли и его математическое описание позволит в существенной степени повысить качество нашей модели.
The study of the process of tumor formation and its mathematical description will significantly improve the quality of our model.
%Кроме того, мы планируем расширить список свободных парметров модели для учета большего числа факторов характеризующих организм человека.
In addition, we plan to expand the list of free model parameters to take into account a larger number of factors characterizing the human body.
%В том числе от каждого отдельно взятого организма будет зависеть сопротивление, оказываемое болезни. 
Including the resistance provided to the disease will depend on each individual organism.
%В уравнении (\ref{eq:equation2.1}) это параметр $\alpha$, который может зависеть от наличия лечения или его отсутствия. 
In the Eq. (\ref{eq:equation2.1}), this is the $\alpha$ parameter, which may depend on the presence of treatment or its absence.
%В дальнейшем возможен учет множества факторов, влияющих на распространения опухоли, таких как возраст организма, неправильное питание, употребление, вредных для здоровья веществ (алкоголь, курение).
In the future, it is possible to take into account many factors that affect the spread of the tumor, such as the age of the body, malnutrition, the use of substances harmful to health (alcohol, smoking).

%Одним из перспективных приложений предложенной модели является создание комбинированных выборок данных для машинного обучения. 
One of the promising applications of the proposed model is the creation of combined data samples for machine learning.
%Такие выборки данных сочетают результаты имитационного моделирования и данные медицинских наблюдений. 
These data samples combine simulation results and medical observations.
%Результаты исследований показывают что применение таких выборок данных позволяет повысить эффективность диагностики онкололгических заболеваний молочных желез \cite{KhoperskovPolyakov}. 
The research results show that the use of such data samples can improve the efficiency of diagnosing breast cancer \cite{KhoperskovPolyakov}.
%Результаты моделирования пространственной динамики роста опухолей могут быть применимы для улучшения качества диагностики по данным маммографии \cite{KARTHIGA2022316}.
The results of modeling the spatial dynamics of tumor growth can be used to improve the quality of diagnosis based on mammography \cite{KARTHIGA2022316}.
%Полученные результаты, однако, позволяют сделать вывод о том, что данная модель удовлетворяет результатам клинических наблюдений и может использоваться для описания развития заболевания. 

The results obtained, however, allow us to conclude that this model satisfies the results of clinical observations and can be used to describe the development of the disease.
%Сравнение показывает хорошее совпадение натурных данных и результатов, полученных при численном анализе данной системы.
The comparison shows a good agreement between the field data and the results obtained from the numerical analysis of this system.

%Основные результаты этой работы: 
The main results of this work:
%1) Предложен метод математического моделирования роста опухоли на основе уравнения диффузии. 

1) A method for mathematical modeling of tumor growth based on the diffusion equation is proposed.
%Этот метод позволяет учитывать пространственную неоднородность среды, что является значимым фактором для подобного рода задач.
This method makes it possible to take into account the spatial inhomogeneity of medium, which is a significant factor for such problems.

%2) Разработано программное обеспечение для вычисления пространственно-временной динамики опухоли. Это программное обеспечение доступно по ссылке ...
2) Software has been developed for calculating spatio-temporal dynamics of tumor. This software is available at the link \url{https://github.com/TenVal/NonLinearCancerModel/tree/findParams3/}

%3) Определена область допустимых значений начальных параметров математической модели методами оптимизации. Диапазон значений для модуля скорости $S$: 1.69 mkm/s $\leq$ $S$ $\leq$ 3.22 mkm/s, а для значений коэффициента распространения опухоли $P$: 1.81 mkm$^2$/s $\leq$ $P$ $\leq$ 3.89 mkm$^2$/s.
3) The area of admissible values of the initial parameters of mathematical model is determined by optimization methods. Range of values for the modulus of speed $S$: 1.69 mkm/s $\leq$ $S$ $\leq$ 3.22 mkm/s, 
%а для значений коэффициента распространения опухоли $P$: 1.81 mkm$^2$/s $\leq$ $P$ $\leq$ 3.89 mkm$^2$/s. 
and for the values of tumor spread coefficient $P$: 1.81 mkm$^2$/s $\leq$ $P$ $\leq$ 3.89 mkm$^2$/s. 

%4) Исследована динамика объема опухоли. Продемонстрирован логистический характер изменения размеров пораженной заболеванием области как для результатов клинических наблюдений, так и для результатов моделирования. Определена относительная погрешность результатов вычислительных экспериментов, которая лежит в пределах от 1.8 процента до 14.6 процента.
4) The dynamics of tumor volume was studied. The logistic nature of change in size of diseased area is demonstrated both for the results of clinical observations and for the results of modeling. The relative error of the results of computational experiments is determined, which lies in the range from 1.8 percent to 14.6 percent.

%5) Исследована пространственная структура смоделированных опухолей. Показана применимость предложенного метода для моделирования неравномерного роста опухоли в зависимости от условий среды.
5) Spatial structure of the simulated tumors was studied. Applicability of the proposed method for modeling uneven tumor growth depending on environmental conditions is shown.

\end{document}